\documentclass[a4paper]{article}
\usepackage[english]{babel}
\usepackage[utf8x]{inputenc}
\usepackage[T1]{fontenc}

\usepackage[a4paper,top=3cm,bottom=2cm,left=3cm,right=3cm,marginparwidth=1.75cm]{geometry}

\usepackage{amsmath}
\usepackage{amssymb}
\usepackage{textcomp}
\usepackage{gensymb}
\usepackage{amsthm}

\usepackage{graphicx}

\usepackage[colorlinks=true, allcolors=blue]{hyperref}
\usepackage{cite}

\usepackage{setspace}
\title{The behavior of capillary suspensions at diverse length scales: from single capillary bridges to bulk}
\author{Sebastian Bindgen$^{\dag,1}$,
Jens Allard$^{\dag,1}$,
Erin Koos$^{1,\ast}$ 
}
 \date{ \small
 $^1$ {KU Leuven, Soft Matter, Rheology and Technology - Department of Chemical Engineering, Celestijnenlaan 200f, 3001 Leuven, Belgium} \\
$^{\dag}$ These authors contributed equally \\
$^{\ast}$ E-mail: erin.koos@kuleuven.be \\}

\begin{document}

\maketitle

\begin{abstract}
\includegraphics[width=0.75\textwidth]{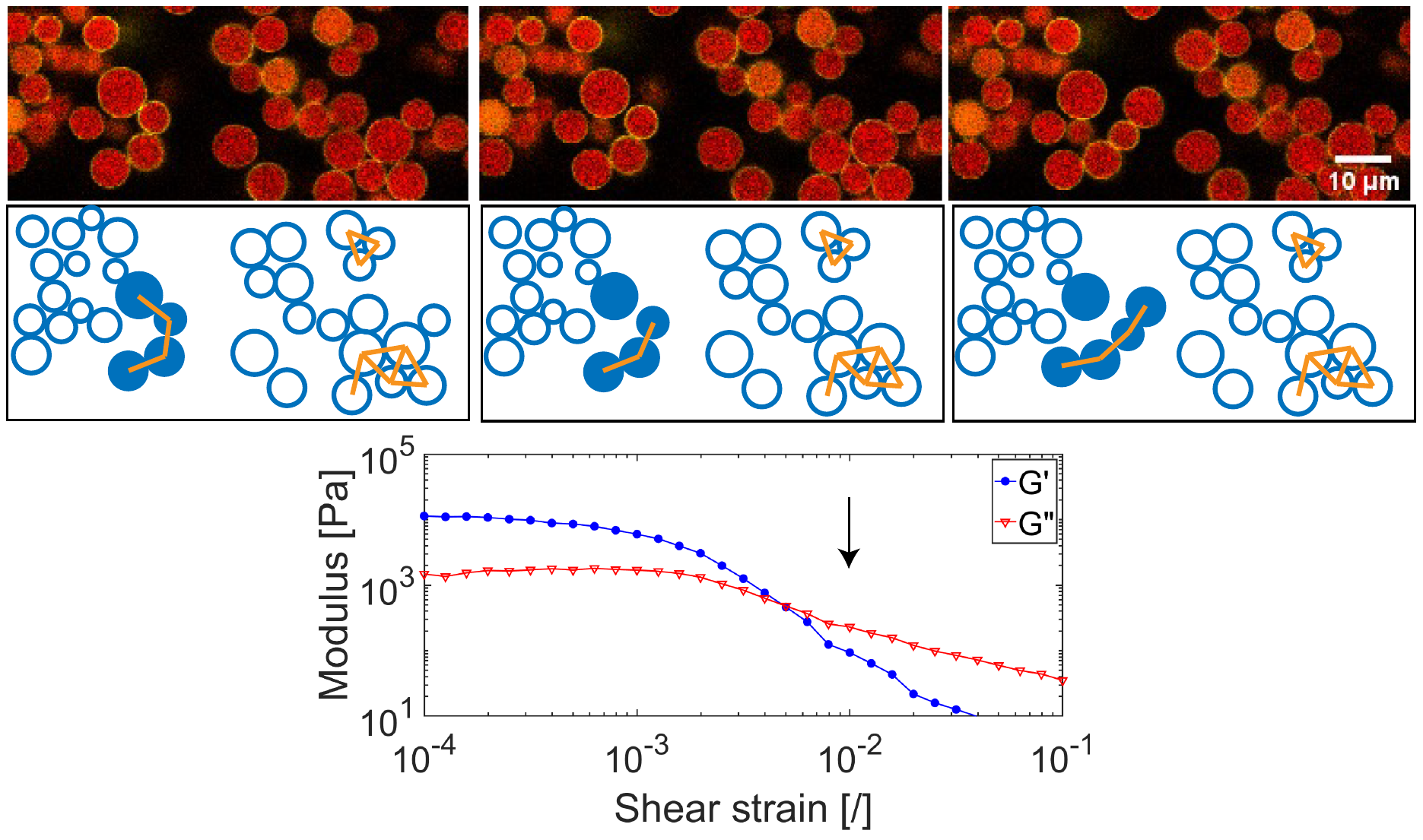}

Liquid-liquid-solid systems are becoming increasingly common in everyday life with many possible applications. Here, we focus on a special case of such liquid-liquid-solid systems, namely, capillary suspensions. These capillary suspensions originate from particles that form a network based on capillary forces and are typically composed of solids in a bulk liquid with an added secondary liquid. The structure of particle networks based on capillary bridges possesses unique properties compared with networks formed via other attractive interactions where these differences are inherently related to the properties of the capillary bridges, such as bridge breaking and coalescence between adjacent bridges. Thus, to tailor the mechanical properties of capillary suspensions to specific requirements, it is important to understand the influences on different length scales ranging from the dynamics of the bridges with varying external stimuli to the often heterogeneous network structure. 
\end{abstract}

\section{Introduction}

Since 2018, Peter Kralchevsky worked on liquid-liquid-solid suspensions with capillary bridges. This research was a natural progression from his work on capillary forces between particles at interfaces, and he was able to provide a unique perspective in this domain. In these capillary suspensions, adding a small amount of secondary fluid to a particle suspension can dramatically alter its mechanical properties.

\begin{figure}[t]
    \centering
    \includegraphics[width=\textwidth]{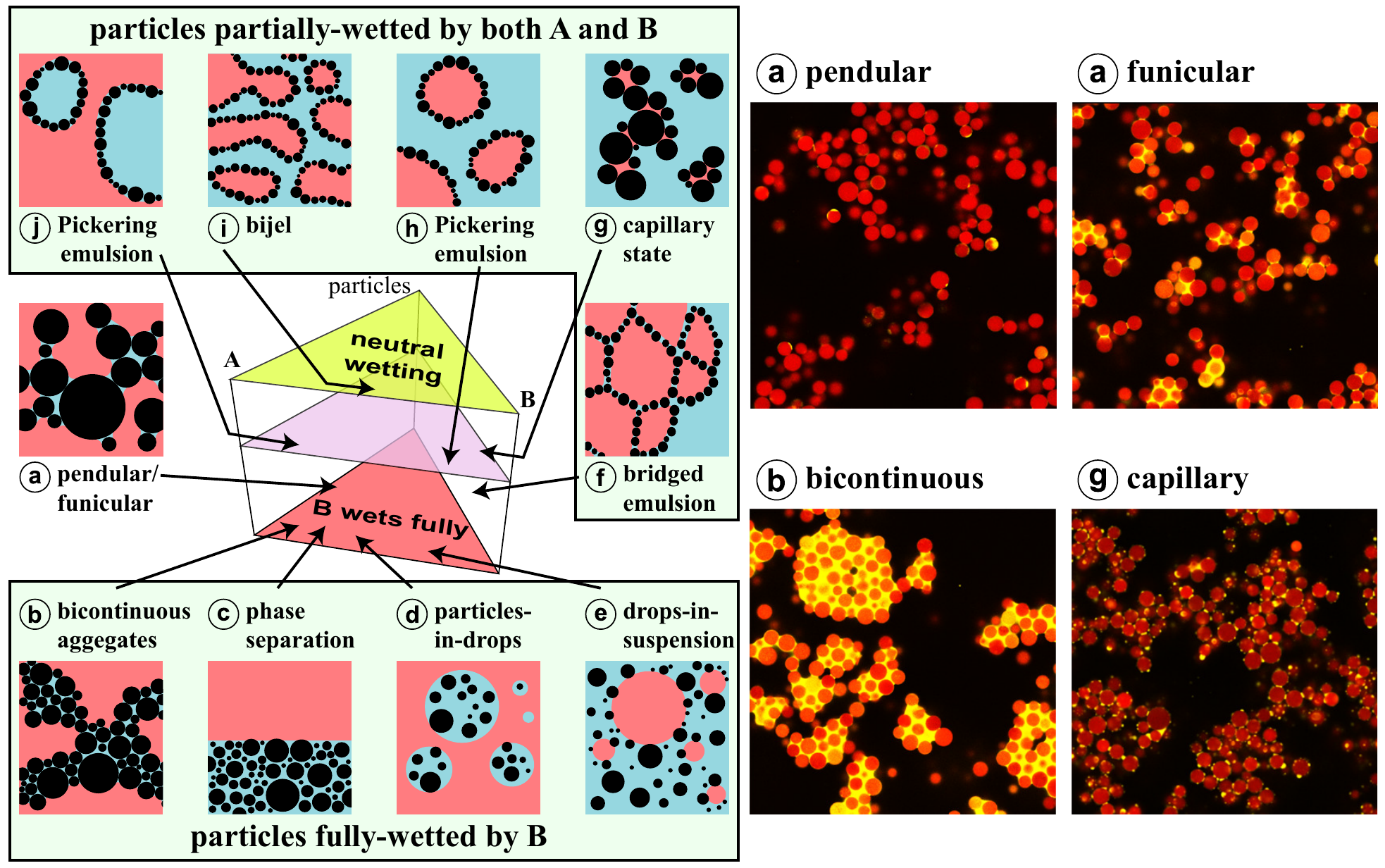}
    \caption{Variety of morphologies observed in liquid-liquid-solid systems with varying composition and particle wettability. State diagram adapted from S.S. Velankar, Soft Matter 43(11), 8393--8403, 2015~\cite{Velankar2015}, republished with permission from the Royal Society of Chemistry; permission conveyed through Copyright Clearance Center, Inc. Images of key states (a,b, g) from S. Bindgen et al., Soft Matter 16(36), 8380--8393~\cite{Bindgen2020}, Copyright 2020 CC-BY-NC. }
    \label{fig:State diagram}
\end{figure}

%%%%
%\subsection{Morphological differences}
Capillary suspensions sit at the intersection of other, more commonly studied materials such as Pickering emulsions, wet granular materials, and even bijels. These are all ternary liquid-liquid-solid systems where the specific morphology depends on the relative volume fractions and the particle wettability. This is mapped by S. Velankar in a three-dimensional state diagram, with the particle wettability on the z-axis, as shown in Figure~\ref{fig:State diagram}~\cite{Velankar2015}. A clear distinction was made between the cases of fully wetted, with a contact angle of 0{\degree}, or partially wetted particles, with a nonzero contact angle, as large differences in microstructure were observed between the two. Later, these state diagrams were experimentally verified using a polyethyleneoxide/polyisobutylene/silica particle model system~\cite{Yang2017,Domenech2017, Yang2018, Amoabeng2017, Amoabeng2020}. Although this diagram is indicative of the regions of existence for the different liquid-liquid-solid structures, there is also an influence of the preparation order and mixing conditions because these structures are not in thermodynamic equilibrium~\cite{Velankar2015, Bossler2017}. Pickering emulsions and capillary suspensions appear in the partially wetted state diagram in which the Pickering emulsions can be found at higher relative amount of dispersed liquid than particles~\cite{Yang2017}.  
In capillary suspensions, the particles are connected by liquid menisci in the pendular state (Figure~\ref{fig:State diagram}a) or are clustered around small droplets in the capillary state (Figure~\ref{fig:State diagram}g), forming a sample-spanning network in either case. The network formation leads to an increase of several orders of magnitude in viscosity and yield stress compared with the binary suspension and prevents particle sedimentation. Capillary suspensions can also be obtained with fully wetting particles, but there is a tendency toward agglomeration and phase separation~\cite{Yang2018}. At a particle loading around 20\%-30\% and both fluids comprising 40\% by volume, some cocontinuous structures were observed. Typically, capillary suspensions are created by adding a secondary liquid, distributed using high shear rates for short periods of time, to a suspension~\cite{Bossler2017}. However, a two-step procedure where the secondary fluid is first emulsified in the bulk fluid and then the particles are added has been used in studies where a uniform microstructure is required~\cite{Domenech2015, Bindgen2020}. Such a procedure will be the focus of further discussion in this review owing to its uniformity and repeatability. 

The stability of capillary suspension networks is subjected to a vast number of influences: the solid load $\phi_\mathrm{solid}$; the amount of added secondary liquid; the surface tension $\Gamma$; the particle radius $a$; and the contact angle $\theta$. In a review, coauthored by Peter Kralchevsky, Danov et al. put forward a general expression of the resulting bulk mechanical properties, for example, the yield stress $\sigma_\mathrm{y}$~\cite{Danov2018}, 
\begin{equation}
\sigma_\mathrm{y} = \frac{2\Gamma}{a} c_\mathrm{g} \phi_\mathrm{solid}^{2/3} f_\mathrm{cap,max}(\tilde{V}, \theta)
%Y (S_i, \phi_p, \Gamma, a, \alpha)
\label{eq:yield_stress}
\end{equation}
this equation arises by considering the capillary force between individual particles and the average number of bridges per unit area. The shape of the bridges, tuning the capillary force between particles, is included in the term $f_\mathrm{cap,max}(\tilde{V}, \theta)$ where $\tilde{V}$ is the volume of the capillary bridge scaled by the volume of the particles, and $\theta$ is the three-phase contact angle of the bridging fluid. The structure of the network is included in the geometric coefficient $c_\mathrm{g}$ where $c_\mathrm{g} \phi_\mathrm{solid}^{2/3}$ is proportional to the number of bridges in the shear plane per unit area~\cite{Danov2018}. Although Equation~\ref{eq:yield_stress} should hold in the limit of higher volume fractions and low secondary fluid volumes, it does not incorporate the more detailed expressions necessary to account for the complex network structure at lower volume fractions or with coalesced bridges. This underlying structure differs in many major points from more classical networks that are, for example, generated by depletion forces. This simplified expression also fails to account for some of the more complex wetting behaviors that can influence both the force and the structure. These two areas will be the focus of the present opinion. 

\begin{figure}[t]
    \centering
    \includegraphics[width=1\textwidth]{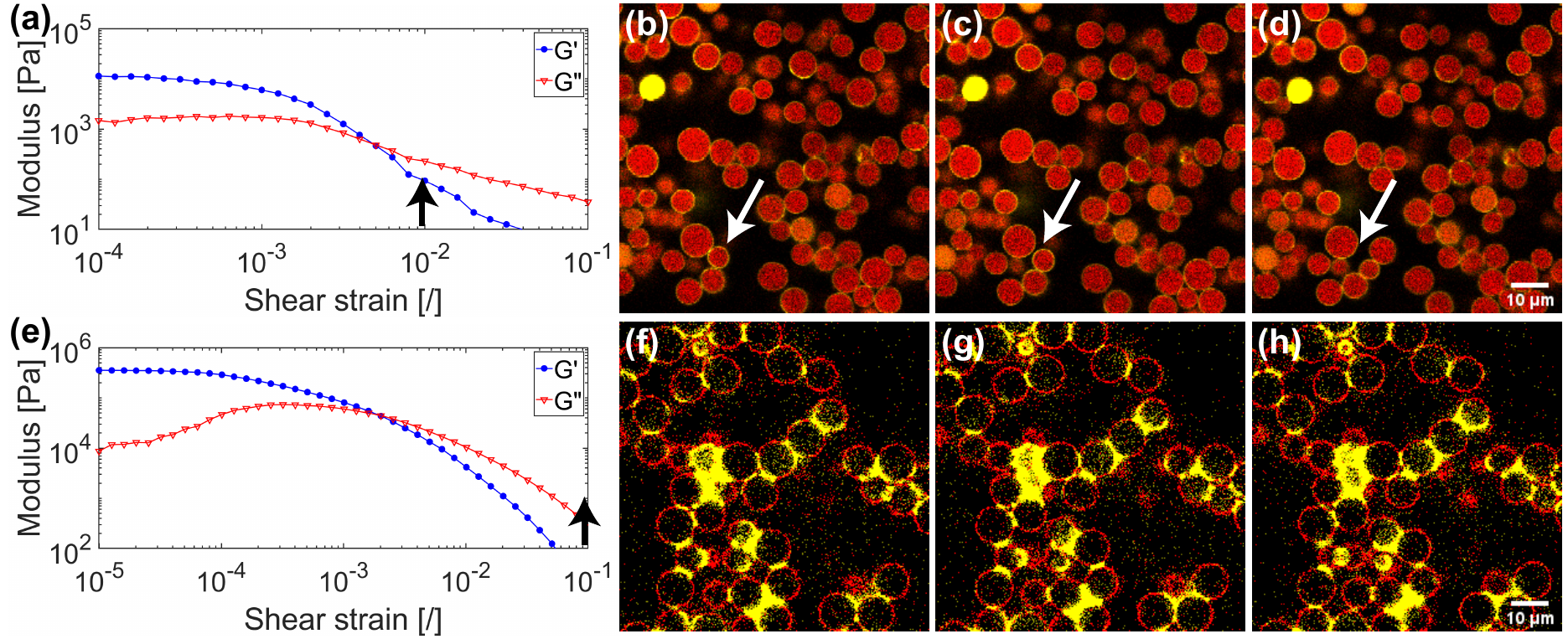}
    \caption{Difference between flexible and rigid capillary suspensions under linear shear. (a, e) Amplitude sweeps and (b-d, f-h) confocal micrographs of capillary suspensions with (a-d) slightly porous and (e-h) nonporous particles. The black arrows on the amplitude sweep indicate the applied shear strain in the confocal micrographs. The white arrows in the top row show a bond-breaking event. Particles are shown as red spheres or circles, the secondary liquid phase is shown in yellow. Composition for the top row: $\phi_\mathrm{solid} = 0.275$ and $\phi_\mathrm{sec} = 0.0275$, for the bottom row: $\phi_\mathrm{solid} = 0.2$ and $\phi_\mathrm{sec} = 0.01$. Time between consecutive frames is 0.14 s with an oscillation frequency of 1 Hz.}
    \label{fig:silica_shearing}
\end{figure}

The particular complexities that arise due to the nature of the capillary force can be demonstrated using a comparison between two capillary suspensions under linear shear, as shown in Figure~\ref{fig:silica_shearing}. The applied oscillatory strain, higher than the crossover strain in both cases, is indicated with the black arrows on the amplitude sweeps (Fig.~\ref{fig:silica_shearing}a, and \ref{fig:silica_shearing}e). In the top row (Fig.~\ref{fig:silica_shearing}a-d), slightly porous particles were used. As a result, some of the secondary liquid is absorbed into the particle pores leaving less liquid available for bridge formation. For this reason, the particles are only connected by small bridges between particle asperities, rather than the larger pendular and coalesced bridges as is the case for the nonporous particles in the bottom row (Fig.~\ref{fig:silica_shearing}e-h). The different wetting behavior and lower amount of secondary liquid result in lower viscoelastic moduli for the porous particles (Fig.~\ref{fig:silica_shearing}a) compared with the nonporous system (Fig.~\ref{fig:silica_shearing}e). This difference is also evident in the observed bond breaking events. In the region examined, particle bonds broke in the sample with weak bridges (Fig.~\ref{fig:silica_shearing}b-d, white arrows), whereas there were no bonds broken for the sample with stronger pendular bridges (Fig.~\ref{fig:silica_shearing}f-h). Both samples are characterized by rigid body movement of the particle clusters, but this is more pronounced for the stiffer network.    

The rigid body motion observed in Fig.~\ref{fig:silica_shearing} is a consequence of the capillary bridges. Particles inside networks formed by capillary forces are explicitly bonded with each other via capillary bridges. This is in stark contrast compared with other particle networks subjected to conservative force fields, for example, depletion interaction, and the additional effects such as contact angle hysteresis, and contact line pinning give rise to changes in the deformation of the networks, such as rigid body movement and, as a consequence, nonsymmetric stress tensors. Therefore, it is essential to investigate the properties of capillary networks on all length scales, making connections between the properties of the bridges and resulting networks. We start by taking a close look at the particle-scale interactions and capillary bonded interactions, discussing implications for the particle networks and describing how these networks can be analyzed. Finally, we discuss how the dynamic wetting of particles as well as other forces can influence both the strength and structure of capillary suspension.

\section{Single capillary bridges in a network}
\label{Section: Single capillary bridges}

\subsection{Capillary force and influence of bridge shape} 

\begin{figure}[t]
    \centering
    \includegraphics[width=1\textwidth]{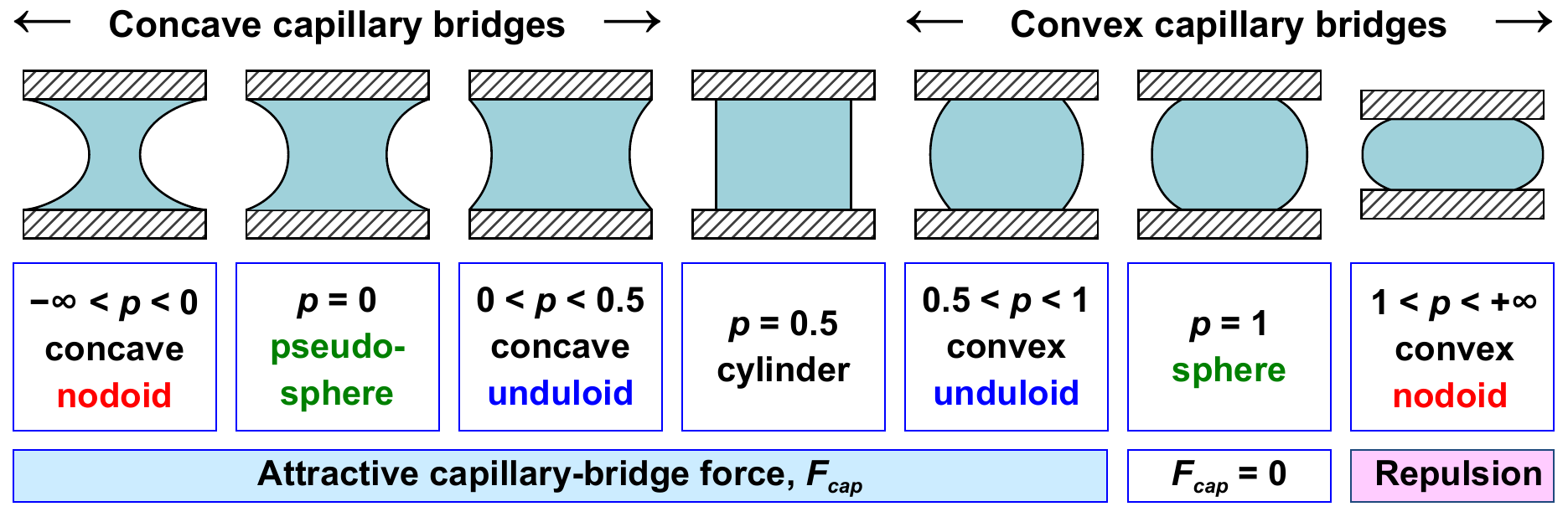}
    \caption{Capillary bridge profiles with increasing dimensionless capillary pressure $p$. Figure reprinted from K.D. Danov et al., Adv. Colloid Interface Sci. 251, 80--96~\cite{Danov2018}, Copyright 2018, with permission from Elsevier.}
    \label{fig:bridge_seq}
    \end{figure}

In the pendular state, particle pairs are connected by capillary bridges. Thus, to understand the macroscopic material properties, we need to examine the properties of single capillary bridges. These capillary bridges minimize the surface energy, which results in bridge shapes with a constant mean curvature, also known as Delaunay surfaces~\cite{Danov2018}. Generally, the capillary force is described by a surface tension contribution and a Laplace pressure contribution. The effect of gravity and other interactions is considered negligible for capillary suspensions, consisting of two liquids and nanoparticles or microparticles. Danov et al. rewrote the capillary bridge force in terms of a dimensionless capillary pressure $p=r_0(P_1-P_2)/(2\Gamma)$~\cite{Danov2018},
    \begin{equation}
    F_{cap} = 2 \pi r_0 \Gamma - (P_1-P_2) \pi r_0^2 = 2 \pi r_0 \Gamma (1-p)
    \label{eq:Danov_capforce}
    \end{equation}
where $\Gamma$ is the surface tension, $r_0$ is the radius of the bridge neck, and $P_1-P_2$ is the pressure difference between the inside and the outside of the bridge. The surface tension term is always attractive. When the pressure inside the bridge is lower than on the outside, corresponding to a negative capillary pressure $p$, the Laplace pressure term is attractive as well. In this case, the bridge has the shape of a concave nodoid, and the capillary bridge force is maximal, as shown in Figure~\ref{fig:bridge_seq}.

With increasing dimensionless capillary pressure, the meniscus profile adopts an increasingly convex shape. When the dimensionless capillary pressure is positive, the Laplace pressure term becomes repulsive but can still be compensated by the attractive surface tension term for values of $p<1$ to yield an overall attractive capillary force. This explains why convex bridges can still exert an attractive force on bonded particles. When the bridge profile is that of a convex sphere at $p=1$, the surface tension and Laplace pressure are equal, resulting in no net capillary force. For $p>1$ the meniscus profile is a convex nodoid with a repulsive capillary force. Because the particle network in capillary suspensions consists of many of these liquid bridges, the force in the bridges on the single-particle length scale is inevitably correlated to the rheological properties on the macroscopic length scale, measured via the yield stress or magnitude of the elastic modulus. Although capillary suspensions can exist in both the pendular state ($\theta < 90\degree$, $p<0.5$) and the capillary state ($\theta > 90\degree$, $p>0.5$), pendular state suspensions with concave bridges are generally stronger than capillary state suspensions with convex bridges~\cite{Bossler2016}. The repulsive bridge formed by a convex nodoid at $p>1$ also implies a limit to the maximum contact angle that the secondary fluid can make against the particle and still form a capillary suspension network.    

Of particular relevance for rheological experiments is the response of the capillary bridges to increasing particle separation. During stretching, the bridge becomes longer and thinner, and the capillary force generally decreases.  This interplay between the contact angle, relative bridge volume, and particle separation has been studied by many researchers~\cite{Willett2000, Butt2009, Megias-Alguacil2009, Xiao2020, Farmer2015} where multiple approaches are possible to calculate the bridge force. The Young-Laplace equation can be integrated from its differential form to obtain the bridge profile. Alternatively, one can assume a certain bridge shape, for example, a Delaunay shape from Figure~\ref{fig:bridge_seq} or a toroid, to directly calculate the force with reasonable precision. A well-known example of such an equation for a toroidal bridge in the limit of small bridges is given by Willet et al.~\cite{Willett2000}:
    \begin{equation}
    F_{cap} = \frac{2 \pi \Gamma a \cos(\theta)}{1 + 1.05 s \sqrt{\frac{a}{V}} + 2.5 s^2 \frac{a}{V}}
    \label{eq:Willet_sep}
    \end{equation}
where $s$ is the particle-particle separation and $V$ the bridge volume. As the third option, the total surface energy can be numerically minimized using software or simulations. Equation~\ref{eq:Willet_sep} was expanded by Alguacil and Gauckler  to describe the regions of existence for concave bridges at larger volumes~\cite{Megias-Alguacil2009}. A minimal separation exists when the bridge volume is large, posing geometrical constraints for the contact angle. At a maximal separation, the bridge breaks. Later, they published a similar existence diagram for convex bridges~\cite{Megias-Alguacil2010}. Convex bridges are possible for contact angles below and above 90\degree, whereas concave bridges only exist for preferentially wetted particles. These convex bridges with contact angles below 90\degree, which exist for high relative volume or low interparticle separation, can transition to a concave shape upon stretching. Using the models of Alguacial and Gauckler and by solving the Laplace-Young equation, Xiao et al. created a model for the capillary force and rupture distance that can take the convex-concave transition into account~\cite{Xiao2020}. This transition can be seen in Figure~\ref{fig:bridge_seq} as the change from convex unduloid to cylindrical bridge and finally to a concave unduloid with increasing interparticle separation. The capillary force initially remains constant, sharply increases at the transition from convex to cylindrical, and then decreases continuously in a similar manner to regular concave bridges. 

A high bridge volume can also cause the bridge to become asymmetric, even without the need for contact angle hysteresis, as was shown by Farmer and Bird using surface energy minimization~\cite{Farmer2015}. If the bridge volume is larger than the volume required to form a spherical bridge and the particles are in contact, the bridge becomes asymmetric with a spherical shape. As mentioned in Figure~\ref{fig:bridge_seq}, these bridges will therefore have no net force acting on the particles. The degree of asymmetry can be quantified by one parameter that is a function of the particle radius, contact angle, and principal curvature along the radial direction. Furthermore, in capillary suspensions, capillary bridges can be formed between spherical particles with a different size or wettability, a problem which was addressed by Sun and Sakai~\cite{Sun2018} and more recently by Wu et al.~\cite{Wu2020}. A short comparison of the different models to calculate capillary forces between two spheres can be found in Yang et al.~\cite{Yang2020}, who also showed that their Lattice-Boltzmann simulations can be effective in calculating the capillary force for the more difficult cases, namely, for higher contact angles and higher relative bridge volumes. They also showed how the attraction from the capillary force can cause the formation of percolating structures in their model.~\cite{Yang2020}  

\subsection{Describing the network structure}

Using the models for single bridge interactions, we can begin to understand both the strength and structure of capillary networks in the pendular state ($\theta<90\degree$) where particle pairs are typically connected by a single pendular bridge (Figure~\ref{fig:State diagram}a). The bridges provide a strong attractive interaction between the particle pairs, similar to other attractive systems. However, the capillary bridges provide an explicit bond between spheres, in contrast to most other attractive forces, for example, van der Waals or depletion interactions, that can be described via conservative force fields. Another difference is that the thermal energy of the particles in a capillary suspension is not able to overcome the attractive potential of the bonds. Despite these differences, we can still use similar tools to describe the properties of such networks of spheres connected by binary contacts. This in turn can help to describe the relationship between the yield stress and material properties, expanding on the relation described by Equation~\ref{eq:yield_stress}.

Because the particles are connected by binary capillary bridges, we can begin by considering the case where the spheres are arranged on a regular lattice. This model provides a good basis to describe the properties of capillary suspensions at high particle load. Direct neighbors are connected by a number of bonds with strength that depends on the amount of added secondary liquid. Removing the restriction of a lattice arrangement results in a disordered structure after relaxation has taken place~\cite{Danov2018}. Rigidity percolation in such disordered systems has been described by Zhang et al.~\cite{Zhang2019} This percolation describes the transition between floppy flocs and rigid networks, and the critical volume fraction where this transition occurs decreases for increasing interaction strength. This leads to the formation of a stress-bearing glass at $\phi_\mathrm{solid} = 0.48$ in their system and, given the strong capillary interaction in capillary suspensions, can help explain the rigid clusters observed at $\phi_\mathrm{solid} = 0.2$ in Figure~\ref{fig:silica_shearing}.

As the particle loading decreases, the arrangement of particles can no longer be described using the abstract lattice models. Starting from the condition of very low particle loading, we can instead describe fractal-like flocs that result from the strong binary interaction between particles. It is important to differentiate between several length scales in such networks. At first, using an analogy between capillary networks in the pendular state and colloidal networks, a fractal dimension that ignores the heterogeneity of the network, was introduced~\cite{Domenech2015}. Based on the scaling law between the yield stress or the storage modulus with the overall solid load, fractal dimensions of $d_f \approx 1.79$ and $d_f \approx 1.98$, respectively, were found. These fractal dimensions are close to the one of wet granular gases. These measurements based on bulk mechanical properties were confirmed and also expanded using different model systems~\cite{Bossler2018}. Here, a difference between dense flocs and a sparse network backbone was introduced. Using data from the yield stress, a fractal dimension of around $d_f \approx 2$, corresponding to the dimension of the backbone, was found. However, using data from oscillatory shear analyzed using the model of Wu and Morbidelli~\cite{Wu2001}, different scaling model fractal dimensions of around $2.57 \leq d_f \leq 2.74$, corresponding to the dimension of the flocs, were found~\cite{Bossler2018}. This indicated that indeed two contributions to the network's strength are important, the interfloc region and the dense flocs themselves. Both dimensions increased with increasing particle size due to a weakening of the capillary force with respect to the particle weight~\cite{Bossler2018}. This was confirmed by confocal microscopy where the fractal dimension was calculated based on the particle number function~\cite{Bossler2018}. 

\begin{figure}[tb]
    \centering
    \includegraphics[width=\textwidth]{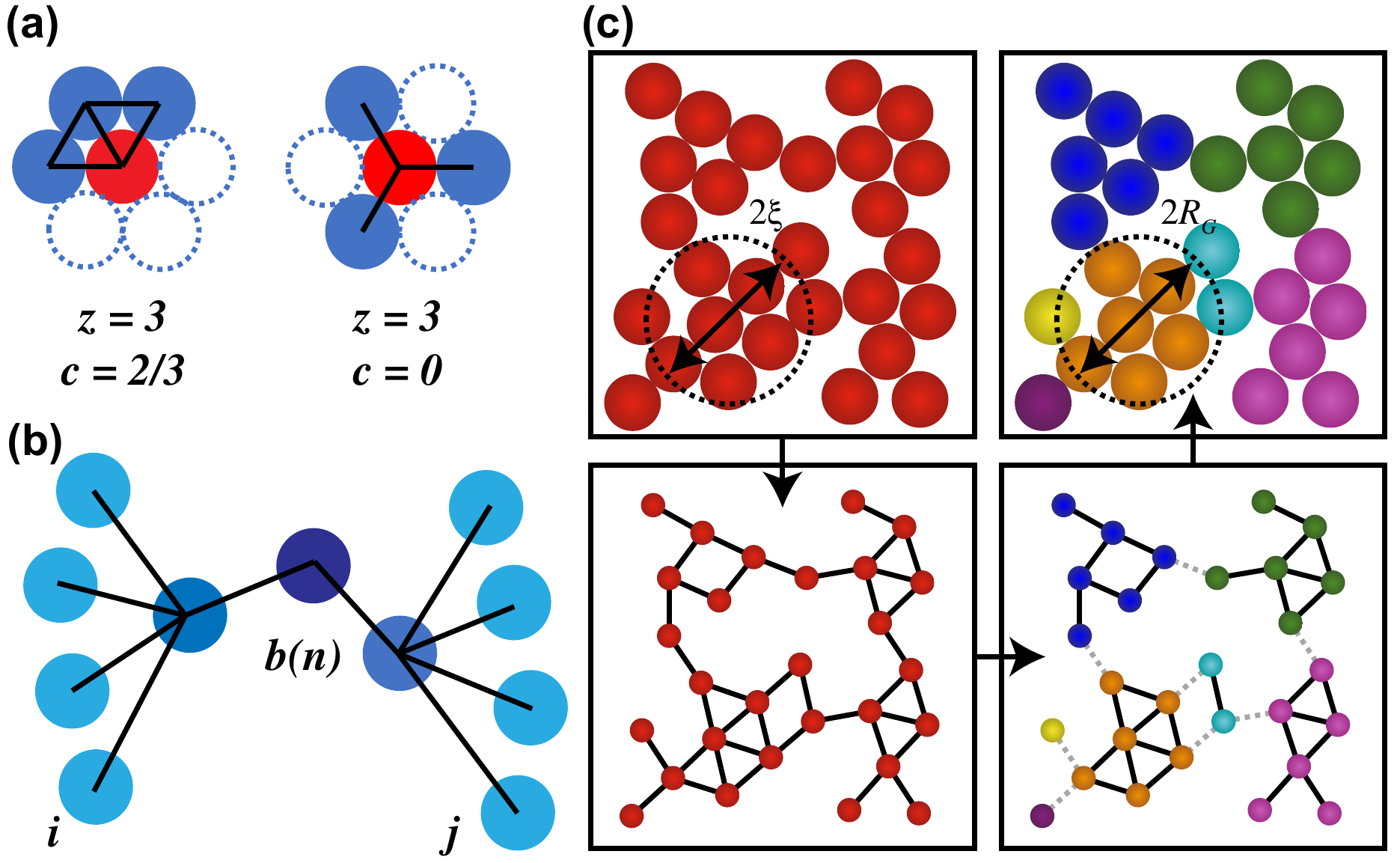}
    \caption{Various graph theory analysis techniques. (a) Clustering coefficient $c$ shown for two configurations (red particle in center) with the same coordination number $z=3$~\cite{Bindgen2020}. (b) Definition of the betweenness centrality $b(n)$ representing the number of pathways through each particle~\cite{Kollmer2019}. (c) Division of a graph into subgraphs for analysis with Cauchy-Born theory~\cite{Whitaker2019}. (a) Republished with permission from S. Bindgen et al., Soft Matter 16(36), 8380--8393~\cite{Bindgen2020}, Copyright 2020 CC-BY-NC. (b) Republished with permission from the Royal Society of Chemistry, from J.E. Kollmer et al., Soft Matter 15(8), 1793--1798, 2019~\cite{Kollmer2019}; permission conveyed through Copyright Clearance Center, Inc. (c) Republished with permission from K.A. Whitaker et al., Nat. Commun. 10, 2237~\cite{Whitaker2019}, Copyright 2019 CC-BY-NC. }
    \label{fig:graphs}
\end{figure}

As shown by Bossler et al., however, the maximum size of the flocs $\xi$ can be small with correlation lengths $10 < \xi/a < 40$ for a solid loading near 25\%~\cite{Bossler2018}. These sizes decrease with increasing solid loading and fraction of secondary fluid. Thus, alternative methods, such as graph theory, must be used to describe the network. In this method, individual particles in a network are interpreted as nodes in a mathematical graph. Neighboring nodes are connected via edges that can be assigned based on a distance or energy criterion. Graph theory has given rise to new measures, such as the clustering coefficient, as shown in Figure~\ref{fig:graphs}a, to describe the structure of particle networks~\cite{Bindgen2020}. This measure was used in combination with the coordination number to develop a state diagram that maps these graph theory measures onto the viscoelastic properties of capillary suspensions. A tight connection between the clustering coefficient and the triangularity of a network is given by the rigidity theory developed by Maxwell. Triangles, or rather particles that are restricted by three bonds, fulfill the isostaticity condition in two-dimensional and are hence stable structures. The more triangles there are in a structure, the more stable it becomes. A similar method using the betweenness centrality (Figure~\ref{fig:graphs}b) was also shown in two-dimensional granular packings to correspond to the observed force network~\cite{Kollmer2019}.

The connection between structure and cluster or floc rigidity is further explored by means of the application of Cauchy-Born theory to colloidal gels where the source of elasticity is described by the $l$-balanced graph partition of these gels~\cite{Whitaker2019}. In this method, the particle networks and, hence, the graphs are cut into subgraphs, as shown in Figure~\ref{fig:graphs}c, in a way that provides the least amount of cuts while retaining $l$ particles in each of the generated clusters.
A strong correlation between the storage modulus and the interparticle attraction potential was obtained for the studied depletion gels~\cite{Whitaker2019}. Furthermore, the data could be mapped onto a state diagram illustrating the connection between the energetic state of the network and the initial volume fraction.

In addition to the number and division of clusters, length scales can be important too. Contributions of particle level potentials need to be distinguished from cluster level contributions to make a suitable prediction for the effective modulus $G$ of a particle network. Johnson et al. highlighted the importance of the gel morphology over the number of bonds as the origin of the gel elasticity using parameters such as the most dominant length scale and the void fraction for Brownian colloidal gels with varying solid volume fraction, aging time, and bond strength~\cite{Johnson2019}. Zaccone et al. define a single length scale $R_0$ that controls the overall modulus $G$ of a gel regardless of the present structural heterogeneities~\cite{Zaccone2009, Zaccone2011, Zaccone2013}. Regardless of the heterogeneity, most of the shear response is determined by intercluster contributions. This means that the storage modulus of capillary suspensions is mostly related to sample-spanning parts of the network, as also shown by the observations using the fractal dimension. The denser flocs mostly redistribute forces into different strands of the backbone~\cite{Bindgen2020, Kollmer2019}. Increasing the amount of secondary fluid only increases the yield stress up to a certain threshold. After that point, the bridges begin to coalesce into dense structures~\cite{Bindgen2020}. These dense clusters grow in size with the addition of secondary fluid but do not actually contribute to the elastic network response. This bridge coalescence is explored further in section \ref{sec:coalesce}.

Recent investigations showed that structural heterogeneity can be connected to the frequency-dependent elasticity~\cite{rocklin2021elasticity}. This was used to derive a phenomenological spring-dashpot model that delivers a good qualitative description based solely on structural information as, for example, delivered by confocal microscope images. Furthermore, the rheology of such colloidal gels can be rate- and history-dependent. This is especially important at low Mason numbers where the ratio of shearing forces to attractive forces is dominating over hydrodynamic effects~\cite{nabizadeh2021life}.

\subsection{Yielding transition and nonlinear properties}

By connecting the structure of particle networks to the bonding mechanism, we can begin to understand the yielding transition in addition to the linear viscoelastic properties. Because the bridges form explicit bonds between particles, we can use their deformation and rupture to predict the transition to yielding. Capillary force-based networks behave quasi-elastically if a shear stress lower than the critical stress is applied. Above this critical stress, which does not have to coincide with the yield stress, the network starts to weaken. On a further increase of the stress amplitude, the system starts to flow. These changes are inherently related to the deformation and stretching of capillary bridges in the pendular state or, for example, the removal of particles from clusters in the capillary state.

The linear viscoelastic regime is very limited for capillary suspensions, typically less than 1\%~\cite{Koos2014}. This can already be deducted from simple approximations of the capillary force as performed by, for example, toroidal bridge approximations. Here, geometrical analysis shows that, even for very small displacements, the force versus distance curves of capillary interactions are highly nonlinear, and both minima and maxima can occur depending on the bridge volumes and contact angles~\cite{Megias-Alguacil2010}. Taking into account that these capillary bridges can also break symmetry~\cite{Farmer2015}, the linear regime is further decreased. Most of the initial decrease in the modulus of a capillary suspension originates from these two effects and not from rupturing bridges. Only later, at deformations higher than the yield point, capillary bridges actually break, and the network breakup starts. A more in-depth study of the yielding behavior revealed a two-step yielding process taking place in capillary suspensions at high deformations in the capillary state. The first part of the yielding process is based on the interaction of separate agglomerates or clusters. The stress increases in the sample when agglomerates collide due to the induced strain~\cite{Ahuja2017}. This effect can be seen if the solid volume fraction is decreased while the amount of secondary fluid is kept constant. The network gets more flexible on a cluster-to-cluster level resulting in an overall flexibility at the network level. 

A process of network breakdown and subsequent build-up can be studied by large amplitude oscillatory shear experiments for capillary suspensions in the pendular state. They revealed two distinct regions based on the applied shear deformation~\cite{Hao2021}: first, the region in which networks are destroyed at low strain amplitudes; and second, the region of network reconstruction at high strain amplitudes. Both processes depend on the volume fraction of particles present and show a clear sol-gel transition between both regions. Such a result is consistent with a structure of dense, more rigid flocs with a number of weaker, floppy connections.

Similar processes have been described for the yielding of attractive colloidal suspensions~\cite{Koumakis2011, Moghimi2020}. Park and Rogers identified an early yielding stage, with rupture of a few bonds, and a late yielding stage with a larger number of bonds rupturing and a breakdown of the network on the cluster length scale~\cite{Park2020}. As for the second part of the yield behavior, there is also a large contribution from the mobility of the clusters that were fluidized during the first part of the yielding process. The more heterogeneous or the more voids there are in the network, the less restrained is the movement of isolated clusters leading to a more pronounced secondary yielding step.

The breakup and restructuring of the clusters can also be seen in steady flow. During flow of pendular state suspensions, isolated clusters, present at high shear rates, follow Jeffery orbits giving rise to a positive normal stress difference $N_1-N_2$. As the shear rate is decreased, these clusters collide and begin to align with the flow to form vorticity rolls, resulting in a negative normal stress difference $N_1-N_2$~\cite{Natalia2018}. The magnitude of this stress difference, however, depends on the ability of the flocs to reorientate in the flow, and the pendular state clusters with small bridges show a more negative $N_1-N_2$ than the more rigid clusters with higher secondary fluid volumes and coalesced bridges. These negative normal stress differences can be used as a tool to generate a fingerprint to distinguish pendular from capillary samples because only negative normal stress differences occur in capillary state samples. The flow-induced arrangement of these clusters can also be used as a method to create pattern formation where the hydrodynamic coupling between flocs and the boundaries determines the periodicity of the log-rolling flocs~\cite{Varga2019}.  
 
\subsection{Bridge coalescence}
\label{sec:coalesce}

The volume of individual capillary bridges between particles will grow with increasing secondary liquid volumes. The growth of the bridges is limited by coalescence between neighboring bridges where the coalescence of the bridges results in the formation of dense clusters inside a capillary network. The growth of clusters depends not only on the amount of secondary liquid present in the suspension but also on geometrical aspects~\cite{Maurath2016}. The resulting network after the transition to dense clusters is usually termed as the funicular state where the network structure is formed not by binary bridges between particles but by bridges between funicular clusters consisting of multiple particles each, as shown in Figure~\ref{fig:State diagram}a. At first sight, one might think that particles are added to these clusters by forming new capillary bridges; however, this effect does not always occur. 
If a secondary fluid with a low contact angle is used, a transition can be observed. Instead of adding particles to existing clusters via new capillary bridges, these bridges begin to merge. This happens because the menisci spread across the particles' surface owing to the low contact angle and subsequently touch each other. As fluids tend to minimize the surface-to-volume ratio, it is energetically favorable that these bridges merge. Hence, three particles connected at first by three isolated bridges can appear to be connected by one central fluid droplet. The resulting structure of the networks is similar to the ones expected from the capillary state.

However, there is an influence that can be seen in the mechanical properties~\cite{Bindgen2020}. The maximum in the dynamic moduli for pendular-funicular state networks with $\theta<90\degree$ appears at a different amount of secondary liquid than for capillary state networks with $\theta>90\degree$. In the former case, the dynamic moduli increase sharply with the addition of secondary liquid until a maximum occurs. After this maximum the moduli decrease again and, hence, the network strength is decreased. This, together with the possible merging of single capillary bridges, is in line with previously mentioned results that state that the strength of the network is mostly influenced by the backbone of the network and less by the amount of bridges inside the clusters. In the case of a capillary state network, the moduli increase until they asymptotically reach the final plateau values. This is an indication that particles merge into already existing clusters with little influence on the interaction strength~\cite{Koos2012}. The behavior of the pendular state suspensions, however, is markedly different. The merging of three single bridges into one trimer is accompanied by possible changes in bridge strength and rupture distance~\cite{Lievano2017,Semprebon2016,Wang2017}. In many configurations, these changes work in opposite direction, for example, a simultaneous increase in capillary force but reduction in rupture distance, which might explain the maximum in capillary force when the funicular state is reached. In experiments with a fully wetted pendular state suspension, Roh et al.~described both a weakening of the storage modulus and a decrease in the critical strain with increasing fractions of secondary liquid~\cite{Roh2017}.  

In both states, regardless of the contact angle of the secondary fluid, the clusters continue to grow with increasing fractions of secondary liquid until they incorporate nearly all of the particles in the system. The secondary liquid has fully coalesced around the particles in a continuous fashion allowing many connections with surrounding particles at once~\cite{Bindgen2020}. The resulting bicontinuous structure still forms a sample-spanning network with a rather thick backbone, and the interstitial regions consist of the bulk liquid devoid of particles~\cite{Domenech2017, Yang2017, Yang2018, Amoabeng2020, Hao2021}, as shown in Figure~\ref{fig:State diagram}b. The existence boundaries for the bicontinuous structures depend on the ratios of the three components~\cite{Yang2017, Domenech2017}, the particle wettability~\cite{Yang2018} and particle surface area~\cite{Amoabeng2020}. The transition to a bicontinuous morphology is noted by a decrease in elasticity compared with the funicular state~\cite{Domenech2017} with fully wetted particles having the highest strength~\cite{Yang2018}. Increasing the aspect ratio promoted the formation of interconnected clusters with more notable two-step yield behavior~\cite{Qiao2019}. Such a bicontinuous network can also be formed via phase separation where the size of the domains depends on the rate of the temperature ramp~\cite{Xi2021, Xi2021b}. With further additions of secondary liquid, this bicontinuous structure finally collapses, and only isolated capillary clusters are left. As the sample-spanning network is not present anymore, also the modulus of the capillary suspension has decreased to almost its initial value~\cite{Bindgen2020}. The system exhibits phase inversion, and the secondary fluid encapsulates the particles in disconnected, round clusters~\cite{Yang2017}.  It is worth remembering, however, that capillary suspensions are not in thermodynamic equilibrium. Such bridge growth and coalescence phenomenon only occur in the case where the mixing is sufficient to break the bridging fluid into drop sizes that are much smaller than the particle size, allowing them to grow and merge based on the wetting properties. This can be further influenced by nonideal wetting and other complications, as discussed in the following sections. 

\section{Wetting dynamics and other complications}

\subsection{Nonideal wetting}
The contact angle between the three phases in liquid-liquid-solid mixtures influences both the morphology and mechanical properties of the system. In the ideal case, that is, no surface roughness or heterogeneity, the three-phase contact angle $\theta$ is determined by the interfacial tensions between the phases, as given by Young's equation:
\begin{equation}
    \cos(\theta) = \frac{\Gamma_\mathrm{solid,bulk}-\Gamma_\mathrm{solid,sec}}{\Gamma_\mathrm{bulk,sec}}
    \label{Eq: Youngs equation}
\end{equation}
where the subscripts \textit{solid}, \textit{bulk}, and \textit{sec} stand for the solid phase, bulk liquid, and secondary liquid, respectively. However, in many systems, particle roughness and other heterogeneous surfaces are present. It is expected that changes to the effective contact angle, contact angle hysteresis, and other complications would affect the properties of both capillary suspensions and other liquid-liquid-solid systems. These effects are expected to play a role in both the dynamic response of these systems and their static properties because the particle networks are in an arrested state, that is, influenced by their history. 

When surface asperities are present on the particle surface, the observed contact angle might be different from the one described by Equation~\ref{Eq: Youngs equation}. In the Wenzel state, the secondary liquid fully wets the particle asperities, and a change in the effective contact angle is given by the increase in the contact line. By contrast, the asperities are filled with air or -- in the case of capillary suspensions -- bulk liquid in the Cassie-Baxter state and heterogeneous wetting occurs. In that case, the contribution from each phase, including any effects caused by the change due to roughness, must be considered. The difference between Cassie-Baxter and Wenzel wetting, in addition to contact line pinning, often contributes to a contact angle hysteresis for rough surfaces. Moreover, the Cassie-Baxter state can transition to the Wenzel state when the droplet impales the asperities, which can be induced by vibrations or evaporation of the droplet~\cite{Papadopoulos2013}.

San-Miguel and Behrens showed that particle roughness, and specifically the difference between Wenzel and Cassie-Baxter wetting, affected the stability of Pickering emulsions~\cite{San-miguel2012}, as characterized using the maximum capillary pressure obtained from centrifugation experiments. Increasing the surface roughness leads to an increase in contact hysteresis and emulsion stability, up to the point where they expect that the wetting regime changed from the Wenzel to the Cassie-Baxter state. A weaker hysteresis corresponds to a lower emulsion stability, even at similar contact angles, owing to a weaker contact line pinning. On top of that, the particle roughness can lead to deformations of the interface in the form of capillary quadrupoles causing a capillary attraction between the particles as another possible stabilization effect. Rough, raspberry-like particles are able to stabilize both oil in water (O/W) and water in oil (W/O) emulsions, depending on the phase in which they were initially dispersed~\cite{Zanini2017}. This variability, caused by the metastable positions of the particles at the interface during mixing, can even be reversed by applied shear. The strong shear is able to overcome the energy barrier associated with metastable pinning and allows for phase inversion~\cite{Zanini2019}. Similarly, a metastable bridging configuration is expected to occur in capillary suspensions. This may either promote or impede to formation of either pendular or capillary state suspensions depending on the mixing order and applied shear history. In a specific case of low relative bridge volume, where multiple small capillary bridges are formed only between particle asperities, that is, each bridge width is of the same order of magnitude as the size of the surface asperities, the particle roughness can increase the total capillary force owing to the migration of each small bridge to the minimum local separation between the particles~\cite{Butler2022}. This migration is only possible if the contact line does not get pinned.   

Bossler and Koos showed that contact angle hysteresis can also affect the structure of capillary suspensions~\cite{Bossler2016}. In their work, porous particles with an apparent contact angle of 115{\degree} still form pendular bridges, which is normally a characteristic of the pendular state for contact angles smaller than 90{\degree}. They reported that the contact angle hysteresis of the particles in this case seemed to shift the pendular to capillary state transition from 90{\degree} to approximately 133{\degree}. Important to note is that the 115{\degree} and 133{\degree} were measured by adsorbing the particles to a flat liquid-liquid interface in a glass microchannel where the particles were initially dispersed in both phases~\cite{Bossler2016}. This difference between the measured contact angle and resulting wetting is clearly visible in the range of apparent contact angles and structures visible in their confocal images~\cite{Bossler2016}. 

To better understand how a range of static contact angles arises in capillary suspensions, it might help to explicitly study the wetting dynamics that arise during the adsorption/desorption of particles, stretching/compression of a liquid bridge, or spreading of the droplet on the particle surface. Generally, two dynamic contact angles are defined: the advancing and the receding contact angle. The advancing contact angle is always higher than the receding angle. For a bridge between two parallel plates, shearing the bridge in the lateral direction also decreases the normal capillary force, as it is maximal when the bridge is in its axisymmetric position and the cosine difference between the contact angles at the left and right contact point is minimal~\cite{Song2021}.

Jiang et al. found a correlation between the spreading force of water droplet wetting hydrophobic polydimethylsiloxane (PDMS) surfaces with pillar and pore topologies, and the advancing contact angle~\cite{Jiang2018}. Both topologies caused Cassie-Baxter state wetting. For the pillar structures, the contact angle increased to $\approx$170{\degree}, regardless of the fraction of pillars, compared with the 118{\degree} for the smooth PDMS surface. For the structures with pores, the contact angle was dependent on the porosity, and the spreading force decreased with increasing contact angle~\cite{Jiang2018}. This distinction between the influence of an apparent roughness between pillars and pores has some clear implications for the presence of a contact angle hysteresis in capillary suspensions. The studies by Bossler and Koos to investigate the structure of capillary suspensions used porous particles~\cite{Bossler2016, Bossler2018}, but most other experiments have been conducted using nonporous particles where asperities are expected. The difference between the two systems and their influence on the contact angle hysteresis might be the focus of future work.   

Because capillary suspension networks are expected to be affected by the contact angle hysteresis, this raises a question about which contact angle (advancing or receding) is most relevant and how the hysteresis influences the capillary force. The receding contact angle is particularly relevant when studying the desorption of particles from an interface or the stretching of a liquid bridge, whereas the advancing contact angle can be important for the adsorption of droplets or particles at an interface and the compression of the liquid bridges. Under shear, some bridges might undergo stretching, whereas others undergo compression depending on their orientation to the shear plane. Shi et al. recorded the dynamic stretching of a liquid bridge between two glass plates at different speeds for water and glycerol bridges~\cite{Shi2018}. They identified four stages per cycle of stretching and compressing the bridge: pinned stretching (I); receding contact line (II); pinned compression (III); and advancing contact line (IV). In stages I and III, the contact radius remained constant, while the contact angle decreased or increased, respectively. In stages II and IV, the contact angle remained constant, whereas the contact radius decreased or increased, respectively. The total dynamic contact angle hysteresis was larger for higher liquid viscosities and higher deformation speeds,  that is, the advancing contact angles were larger, and the receding contact angles were smaller. This showed that the contact angle hysteresis, in addition to the capillary force, is influenced by the viscous force. The total force increased to a maximum at the end of the pinned stretching stage, after which it decreased to a minimum at the end of the pinned compression stage. 

In 2016, Anachkov et al.~used colloidal probe atomic force microscopy (AFM) to investigate the dynamic contact angle of single particles at an air-liquid or liquid-liquid interface~\cite{Anachkov2016}. When pulling a particle from the interface, a liquid meniscus forms. They managed to calculate advancing and receding contact angles from the force-displacement curves when the particle approached and subsequently was retracted from the interface. The experiment could be performed in a force-controlled or displacement-controlled manner. In the former case, the particle detached from the interface when the maximum force is obtained. Using displacement control, the liquid bridge was stretched a little bit further than the point of maximum force. Thus, the maximum displacement at which the contact line detached ($D_0$) and the displacement at maximum force ($D_{max}$) did not coincide. Two years later, two similar studies with colloidal probe AFM on particles at interfaces were made by Schellenberger et al.~\cite{Schellenberger2018} and Zanini et al.~\cite{Zanini2018}.  In the study by Schellenberger et al., a colloidal probe setup was combined using a confocal microscope to image the meniscus profile during the approach and retraction process~\cite{Schellenberger2018}. The contact line was pinned for most of the retraction process, whereas it started to slide only at the end of the process, rather than the limiting cases of either pinned or sliding contact line. This situation can be compared with the pinned stretching and receding contact line described by the aforementioned work of Shi et al.~\cite{Shi2018}. As in the experiments of Shi, Schellenberger found that the maximum force occurred at the end of the pinned stretching stage. The conclusion of this study was that the receding contact angle has to be used to calculate maximum capillary force for particle detachment from an interface. If we link these studies to capillary suspensions, where the yield stress is correlated with the extension of the bridges, it might be more accurate to use the receding contact angle in models like the one from Danov~\cite{Danov2018}. Simply using the receding contact angle is not valid for every system, however. Using differently functionalized rough particles, Zanini et al.~\cite{Zanini2018} found that contact line sliding occurred for smooth particles, as shown by the mismatch between $D_0$ and $D_{max}$. For rough particles, these values were closer to being equal, indicating particle detachment via a pinned contact line. Furthermore, smooth but chemically heterogeneous particles also exhibited contact line pinning similar to rough particles. The observation of surface corrugations of the interface around the fluoro-functionalized particles further confirmed that the chemically heterogeneous patches had a similar effect as particle roughness.   

\begin{figure}
    \centering
    \includegraphics[width=1\textwidth]{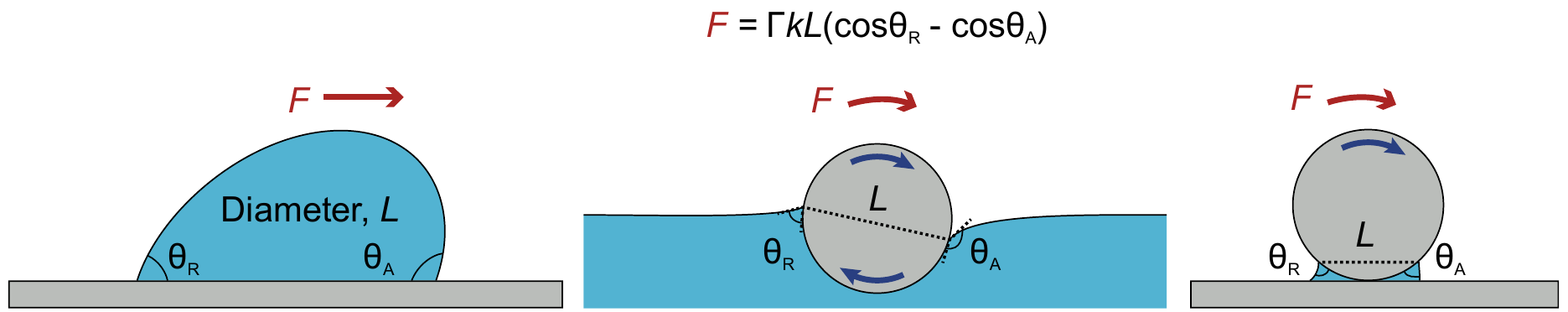}
    \caption{Comparison of the three essential movements present in capillary suspensions. Each of them results in the formation of an advancing and a receding contact angle: (left) sliding of a droplet across a flat surface, (center) rotation of a sphere at an interface between two phases, and (right) start-up of a sphere rolling across a flat surface attached via a capillary bridge. Adapted with permission from A. Naga et al., Langmuir 37(24), 7457--7463~\cite{Naga2021}, Copyright 2021 CC-BY-NC. }
    \label{fig:bridgesrotation}
\end{figure}

Contact dynamic effects, such as contact angle hysteresis or contact line pinning, are certainly important when capillary bonds are stretched or compressed but can also arise under rotation, as shown in Figure~\ref{fig:bridgesrotation}~\cite{Naga2021}.
Naga et al. recently showed that a particle situated at an interface between two phases can experience similar effects as a liquid droplet sliding across a surface or a sphere rotating above a surface. The resulting force or torque forms a barrier similar to an activation energy that needs to be overcome to reach a state of unhindered sliding or rotation. This capillary torque is given by
\begin{equation}
    M = \Gamma RLk(\cos \theta_R - \cos \theta_A)
\end{equation}
with the parameter $k$ as geometrical constant of $k=24/\pi^3$.
Particles connected by a capillary bridge are not able to freely rotate around one another, and the particles within a capillary network will transfer that torque to neighboring particles in the network. Only by overcoming the threshold torque, particles are able to rotate, and the contact line moves along the surface. This also had an impact on the force required to detach rotating particles from a liquid interface, with a decrease of around 25\% compared with nonrotating particles.~\cite{naga2021force}
Similarly, a threshold torque for the yielding of silica particle rods in saline water was discovered using optical tweezer experiments~\cite{Bonacci2021}. As the capillary force should create even stronger particle contacts, this hindrance is likely the cause of rigid body movement that occurs when deforming capillary networks, as shown in Figure~\ref{fig:silica_shearing}. 

Gelling agents offer another route to influence the yield stress of capillary suspensions. Das et al.~found this effect by using a watery secondary phase together with either methylcellulose or agarose~\cite{Das2017}. They created thermo-responsive capillary suspensions which either increased the yield stress with increasing temperature (case of methylcellulose) or decreased the yield stress with increasing temperature (case of agarose). Although the addition of such gelling agents is mostly restricted by the composition and hence the solubility of these agents in the secondary phase, it offers an interesting way to control final material properties. Such effects of gelling agents have been used in the creation of sintered ceramics with capillary suspensions as precursor~\cite{Yang2020yield}. Precise control of the added gelling agent (sodium alginate and excess Ca-ions) allowed the final pore size and pore size distribution to be tuned~\cite{Yang2020yield}. One effect that contributes to this is the enhanced stiffness of gelled capillary bridges. On shearing, the capillary suspension breaks down into bigger aggregates rather than single particles. 

\subsection{Other interactions} 

When properly mixing the three phases, a capillary suspension forms almost instantly with a yield stress described by Equation~\ref{eq:yield_stress}. In this section, we discuss two phenomena, mixing and rewetting as well as electrostatic interactions, which can have a large influence on some systems but are not taken into account when using Equation~\ref{eq:yield_stress}.

In theory, the same liquid/liquid/solid system should be able to create both a pendular and a capillary state capillary suspension by just changing the ratio of the two liquids. In practice, however, we see that each material combination has a preference toward one of the two states, making it difficult to prepare the other one. Although the secondary fluid can be used to modify particle wettability in situ~\cite{Dunstan2018}, there still appears to be other interactions at play.
Bossler and Koos showed that for their viscosity-matched system of glycerol and silicone oil, only the pendular state could be achieved~\cite{Bossler2017}. Instead of a capillary state capillary suspension, an O/W Pickering emulsion was obtained. The morphological difference might seem subtle, in the sense that particles are adsorbed on droplets in both systems. However, the particles are adsorbed on droplets that are much larger than the particle size in a Pickering emulsion and, at the used volume fractions, these droplets do not form a sample-spanning network, thereby not increasing the sample strength~\cite{Kaganyuk2019}. Unfortunately, the stabilization of Pickering droplets appears to be favored for particles that are preferentially wetted by the bulk liquid, as is the case when trying to prepare a capillary state capillary suspension. This favorability, however, is influenced by the physical properties of the liquids and particles. For instance, the viscosity ratio plays a pivotal role in the droplet breakup behavior during the preparation of a capillary suspension, favoring the formation of smaller droplets if the bulk liquid has the higher viscosity~\cite{Grace1982}. The problems caused by droplet breakup, particularly in the presence of the particles, was one reason that the secondary fluid was emulsified before adding the particles in the polymer melt system of Domenech and Velankar~\cite{Domenech2015}~\cite{Domenech2014}~\cite{Domenech2017} as well as the small sample volumes used for confocal microscopy~\cite{Bossler2016, Bindgen2020}. However, the breakup behavior of both liquids should be the same in a viscosity-matched system, and hence other forces or interactions dominate the observed behavior in the work of Bossler and Koos, preventing the formation of the capillary state suspension in that system~\cite{Bossler2017}. 

\begin{figure}[tb]
    \centering
    \includegraphics[width=\textwidth]{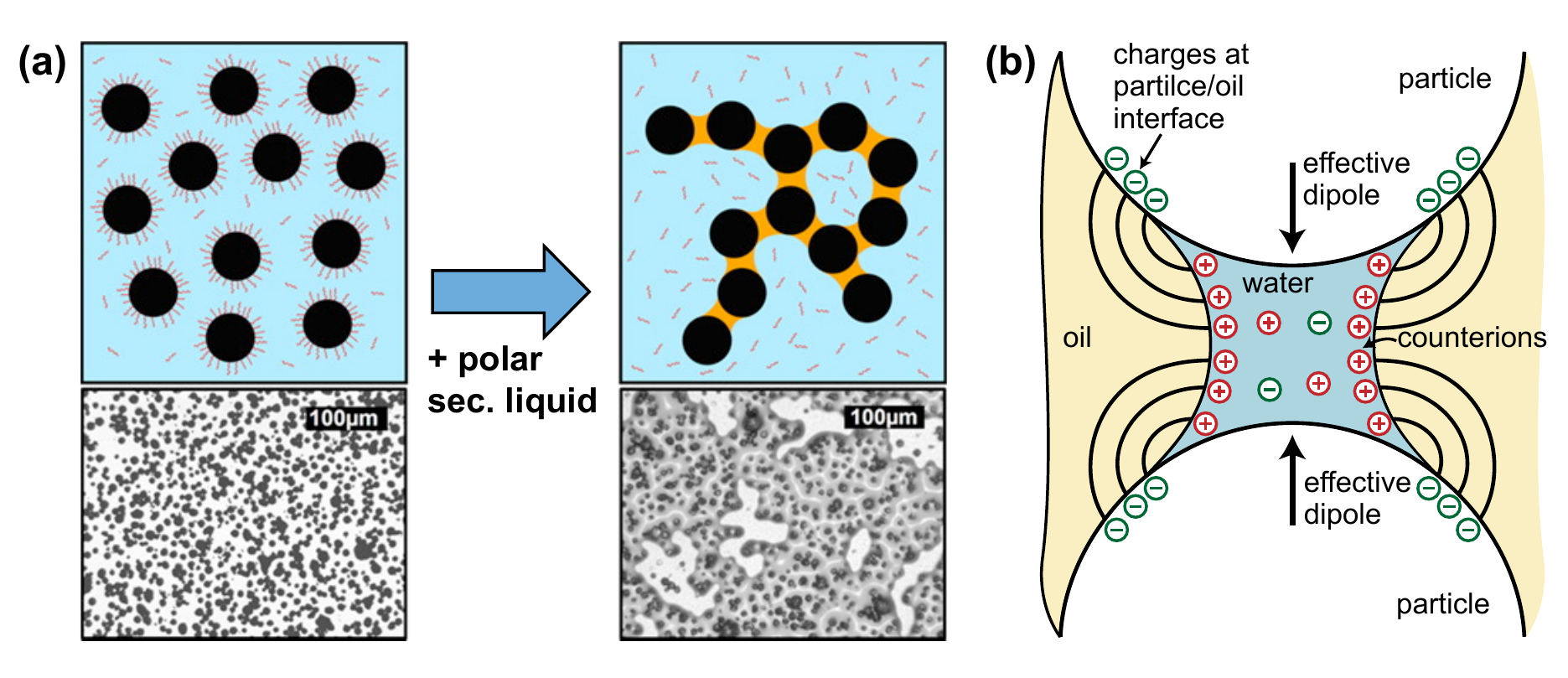}
    \caption{Noncapillary interactions are often neglected, but can have profound effects in the structure and strength of the effective attractive interactions. (a) Preferentially-wetting liquids are able to displace polymer layers from the particle surface and form a capillary network~\cite{Bitsch2016}, (b) electrostatic repulsion across the oil phase, caused by charges at the oil-particle interface, counteracts the capillary force~\cite{Georgiev2018}. (a) Reprinted with permission from B. Bitsch et al., Langmuir 32(6), 1440--1449~\cite{Bitsch2016}. Copyright 2016 American Chemical Society. (b) Reprinted with permission from M.T. Georgiev et al., J. Colloid Interface Sci. 513, 515--526~\cite{Georgiev2018}, Copyright 2018, with permission from Elsevier.}
    \label{fig:Bitsch_Georgiev}
\end{figure}

The favorability in producing either a pendular state or capillary state suspensions can also be influenced by the dynamic wetting behavior of the particles. To generate a particle network, the particles in a suspension have to collide with and be wetted by secondary liquid droplets. If the particles cannot be wetted by the secondary liquid droplets, it might explain why either the pendular or capillary state is preferred. Some very interesting results on this subject were obtained by Bitsch et al.~\cite{Bitsch2016} using graphite particles suspended in different aqueous polymer suspensions. In the case of a nonadsorbing polymer, both the pendular and capillary state could be obtained by using polar ($\theta<90\degree$) and apolar ($\theta>90\degree$) secondary liquids, respectively. In contrast, when the polymer adsorbed onto the particle surface, only the polar liquids were able to form a particle network, as shown in Figure~\ref{fig:Bitsch_Georgiev}a. They showed that the polar liquids displace the adsorbed polymer layers, while the apolar liquids are unable to do so. Even though the three-phase contact angle for apolar liquids remained the same for particles with adsorbing and nonadsorbing polymers, the interactions between the adsorbed polymer layers and secondary apolar liquid droplets prevented the formation of a particle network, possibly due to steric repulsion by the polymers on the particle surface. It is already known that the wetting dynamics play a role in other three-phase systems~\cite{Velankar2015}. In Pickering emulsions, for example, the droplet size and phase inversion composition are affected by the mixing order, namely, in which phase the particles are first dispersed~\cite{Binks2000}. The Pickering emulsions were more stable, indicated by a smaller droplet size, when the particles were predispersed in the continuous phase rather than the droplet phase. In most preparation methods for capillary suspensions, the particles are predispersed in the bulk phase, which might explain why the secondary liquid droplets might not be able to wet the particles in some cases. 

A second problem might be the presence of electrostatic interactions. These interactions are present for particles adsorbed at a liquid-liquid interface~\cite{Danov2004}, and thus might also affect interactions across one or multiple interfaces, that is, oil-water, particle-oil, or particle-water, in liquid-liquid-solid systems. Georgiev et al. studied the difference between oil-continuous and water-continuous pendular state capillary suspensions using the same liquids and hydrophilic or hydrophobic silica particles, respectively~\cite{Georgiev2018}. For water-continuous suspensions, 0.5~M of NaCl was added to the water phase to suppress any electrostatic repulsion between the oil droplets and the silica particles during sample preparation, because this repulsion hindered the formation of capillary bridges. For their soybean oil-water system, the oil-continuous suspension had a lower yield stress than the water-continuous system, even though the contact angles were similar and the higher oil viscosity would favor better droplet breakup for the oil-continuous system. The weaker yield stress was explained by an electrostatic repulsion across the oil phase between the particles that counteracts the capillary force, as shown in Figure~\ref{fig:Bitsch_Georgiev}b. Owing to the negligible charge density in the oil phase, two effective dipoles are created. When the force contribution from these dipoles was taken into account, the authors were able to calculate the charge density at the particle-oil interface from the yield stress data using the model derived by Danov et al.~\cite{Danov2018}. The electrostatic repulsion becomes relatively more important for small bridge volumes. In the other system tested in their paper, hexadecane-water, the effect of the electrostatic repulsion was negated by the higher interfacial tension between the liquids. Nonetheless, Georgiev's findings convincingly point to the fact that the electrostatic interactions in single bridges can have an important effect on the bulk rheological properties.     

\section{Conclusions and outlook}

Capillary suspensions are liquid-liquid-solid systems in which particles are connected by liquid menisci forming a sample-spanning network. The particles are explicitly bonded by capillary bridges, which evapos different properties compared with other attractive networks. The capillary force of each bridge depends on the meniscus shape, which in turn is controlled by many parameters such as bridge volume, three-phase contact angle, and particle size and shape. Calculating the capillary bridge force with high accuracy, particularly in response to the external displacement, can still prove to be a challenging task, particularly when the relative bridge volume is high or the bridge symmetry is broken. In the present review, we described how the changing wetting behavior can lead to different morphologies and mechanical material response.

The capillary suspension network cannot be described by conservative force fields like other attractive networks, and the capillary bonds are much stronger than the thermal energy of the particles. At high solid loading, when the structure is more uniform, lattice models have successfully been used to estimate rheological properties~\cite{Danov2018}. At low solid loading, when fractal-like flocs are present, the elastic properties of the network can be calculated by rheological scaling laws, or alternatively the network can be analyzed using concepts from graph theory. To accurately predict rheological properties like the yield stress, the important length scales have to be identified. For yielding of capillary suspensions, a two-step process is often assumed in which first clusters become more mobile, for example, owing to a small amount of intercluster bonds breaking, after which the structure breaks down more. With increasing amounts of secondary liquid, capillary bridges between particles are formed, and the existing bridges grow. This growth leads to coalescence with neighboring bridges, marking the transition from the pendular to the funicular state where the bulk mechanical properties peak. A further increase in secondary liquid leads to bicontinuous aggregate structures which eventually also break down, and the sample-spanning network is lost. 

The three-phase contact angle is a key feature which affects both the morphology and strength of capillary suspensions. Contact angle hysteresis is expected to affect capillary suspensions both statically, via the apparent contact angle of the bridges, and dynamically, when the suspension is sheared and bridges are stretched and compressed. Finally, we described rewetting and electrostatic interactions as reasons why some systems might not be able to form pendular or capillary state capillary suspensions. Besides the relative volume fractions and particle wettability, the final morphology of these ternary systems is also influenced by preparation method. Recently, alternative mixing strategies for capillary suspensions have been discussed~\cite{Fischer2021c}. Here, capillary suspensions are created with partially miscible liquids, however, without undergoing temperature changes that would, for example, result in phase transitions. A similar effect was also demonstrated for an electrorheological suspension in a mixture of silicone oil and alkane~\cite{Liang2020}. Simulations have also been conducted on such binary liquids under shear showing that while the aggregation size changes with shear rate, the concentration profile around a particle is insensitive to shear~\cite{Barbot2017}

Although we are beginning to understand the role of bridge coalescence in the strength and structures of capillary structures, the reverse situation caused by evaporation of the bridging fluid is not always marked by the same transition from a dense cluster to sparser bridged network structure. While this can be advantageous for the formation of porous supraparticles~\cite{Tan2019} or in the three-dimensional printing of structured porous materials~\cite{Maurath2017, Roh2017, Ding2020, Weiss2020}, the changes due to evaporation should be better understood. The bridges do seem to persist for long times, even if they have a higher vapor pressure than the bulk liquid~\cite{Fischer2021b}, leading to a reduction in cracking~\cite{Schneider2017, Fischer2021, Gastol2021, Park2021}, but the link between the drying conditions and formulation should be better understood. Furthermore, the influence of the bridges on particle motion and segregation of small particles, for example, binder, should be explored.  
Finally, the role of capillary forces caused by the addition of a binding fluid at interfaces is important to understand. The interactions between particles can cause, for instance, the transition between adsorption of soft microgel particles at an interface with crystalline arrangement or the desorption of the particles from the interface for the collapsed particles at higher temperatures~\cite{Wang2019}. The adsorption of binder particles to the liquid-liquid interface is also seen in capillary suspensions where the smaller binder particles can even weaken the network at high volumes or long times~\cite{Park2019}.
The adsorption of the particles to the interface is enhanced in the presence of a capillary interaction. Here, quaternary air/oil/particles/water capillary foams exhibit increased strength and long lifetimes~\cite{Okesanjo2020, Behrens2020}. This is caused by the synergistic action of the particles and the added fluid at the bubble interface~\cite{Zhang2014, Zhang2015, Zhang2017}. The oil and particle coated droplets are incorporated into the gel network of a capillary suspension forming structures that can be easily solidified~\cite{Zhang2014, Zhang2015}. Capillary foams show formation boundaries that are more complex than either Pickering emulsions or capillary suspensions, owing in part to the complex interactions, and their formation seems to be favored in cases of near-neutral wetting oil with high effective oil spreading coefficient~\cite{Zhang2017, Behrens2020}. Although the capillary foams can be created using different mixing orders~\cite{Zhang2015}, wetting dynamics may still play a role in their formation and destabilization in addition to the rheological response. 

\section*{Acknowledgements}
The authors would like to thank financial support from the Research Foundation Flanders (FWO) Odysseus Program (grant agreement no. G0H9518N) and International Fine Particle Research Institute (IFPRI).

% \bibliographystyle{elsarticle-num} 
% \bibliography{bibliography}

\end{document}